# Searching for the flavon in the diphoton channel at future super hadron colliders

M. A. Arroyo-Ureña,[1,2,*] Diego Carreño,[1,2,†] and T. A. Valencia-Pérez[3,‡]

[1]*Facultad de Ciencias Físico-Matemáticas, Benemérita Universidad Autónoma de Puebla,*
*C.P. 72570, Puebla, Puebla, México*
[2]*Centro Interdisciplinario de Investigación y Enseñanza de la Ciencia,*
*Benemérita Universidad Autónoma de Puebla, C.P. 72570, Puebla, Puebla, México*
[3]*Instituto de Física, Universidad Nacional Autónoma de México, C.P. 01000, CDMX, México*

We study the production of a *CP*-even flavon $H_F$ in proton-proton collisions and the prospects for its detection via the diphoton channel at future super hadron colliders, i.e., $pp \to H_F \to \gamma\gamma$. The theoretical framework adopted is a model that invokes the Froggatt-Nielsen mechanism with an Abelian flavor symmetry, which includes a Higgs doublet and a complex singlet. We confront the free parameters of the model against theoretical and experimental constraints to find the allowed parameter space, which is then used to evaluate the production cross section of the flavon and the branching ratio of its decay into two photons. We find promising results based on specific benchmark points, achieving *signal significances* at the level of $5\sigma$ for flavon masses in the interval $200 \lesssim M_{H_F} \lesssim 450$ and integrated luminosities in the range 5–12 ab$^{-1}$ at the future High-Energy LHC. On the other hand, the Future Circular Collider-hadron-hadron could probe masses up to 1 TeV if it reaches an integrated luminosity of at least 2 ab$^{-1}$.

## I. INTRODUCTION

The Standard Model (SM) of particle physics [1–4] has been tested over the last five decades and has been shown to successfully describe elementary particle interactions, including the mechanism responsible for breaking electroweak symmetry [5–10]. This mechanism gives mass to the massive particles of the SM and predicts the existence of the Higgs boson (*h*). It was verified at the LHC in 2012 [11,12]. One of the studied channels focused on the process $h \to \gamma\gamma$, and was shown to be compatible with the prediction of the SM, being the last elementary particle discovered within the theoretical framework of the SM, which proved to be a successful theory. However, it is well known that the SM cannot explain phenomena such as the nature of dark matter, the mass hierarchy problem, and the flavor problem, among others. There are a plethora of extensions that attempt to explain one or more of the open questions [13–22]. One possibility is the Froggatt-Nielsen mechanism [23], which tries to explain the hierarchy of fermion masses. This mechanism assumes that above some energy scale $\Lambda$ there is a symmetry, perhaps of Abelian type $U(1)_F$ (with the SM fermions being charged under it), which prohibits the emergence of Yukawa couplings at the renormalizable level. Yukawa matrices can arise through nonrenormalizable operators, though. The Froggatt-Nielsen model (FNSM) predicts the so-called *flavon* ($H_F$), which has been studied in its "golden channels," namely, $H_F \to ZZ$, $H_F \to WW$, and $H_F \to hh$ [24–26]. However, the diphoton channel has not yet been explored in a comprehensive way. This may be due to the suppression of the $H_F \to \gamma\gamma$ decay rate, mainly due to the couplings involved in the loop that induces this decay, namely, $g_{H_F t\bar{t}} \sim 0.01 m_t/v$ and $g_{H_F W^- W^+} \sim 0.01 g_{hW^- W^+}$. Nevertheless, it has the advantages of good resolution on the flavon mass $M_{H_F}$ and small QCD backgrounds.

In this paper, we are interested in the possible detection of the flavon via the diphoton channel in the theoretical framework of the FNSM. In particular, we study a special scenario in which the flavon is a quasifermionphobic particle, which couples mainly to gauge bosons and an *hh* pair. The study is focused on future proton-proton *pp* colliders:

(i) High-Luminosity Large Hadron Collider (HL-LHC) [27]. The HL-LHC is a new stage of the LHC starting about 2026 at center-of-mass energy of 14 TeV. The upgrade aims at increasing the integrated luminosity by a factor of 10 (3 ab$^{-1}$, year 2035) with respect to the final stage of the LHC (300 fb$^{-1}$).

---

[*]Contact author: marco.arroyo@fcfm.buap.mx
[†]Contact author: diego.carrenod@alumno.buap.mx
[‡]Contact author: tvalencia@fisica.unam.mx

(ii) High-Energy Large Hadron Collider (HE-LHC) [28]. The HE-LHC is a possible future project at CERN. The HE-LHC will be a 27 TeV $pp$ collider being developed for the 100 TeV Future Circular Collider (FCC). This project is designed to reach up to 12 ab$^{-1}$, which opens a large window for new physics research.

(iii) Future Circular Collider-hadron-hadron (FCC-hh) [29]. The FCC-hh is another possible future 100 TeV $pp$ hadron collider that may be able to discover rare processes, new interactions up to masses of around 30 TeV and search for a possible substructure of the quarks. Because the high energy and collision rate, up to $10^5$ flavons may be produced. The FCC-hh will reach up to an integrated luminosity of 30 ab$^{-1}$ in its final stage.

This work is structured as follows: in Sec. II, we conduct a comprehensive review of the FNSM. Experimental and theoretical constraints on the model parameter space are also included. Section III focuses on taking advantage of the insights gained from previous section, performing a computational analysis of the proposed signal and its SM background processes. Finally, the conclusions are presented in Sec. IV.

## II. THEORETICAL FRAMEWORK

In this section, we present the relevant theoretical aspects of the FNSM. In Refs. [24,25,30–32], a deep theoretical analysis of the model and a study on the model parameter space are reported.

### A. Scalar sector

The scalar sector includes one singlet complex scalar $S_F$ to the SM. In the unitary gauge, the SM Higgs doublet $\Phi$ and $S_F$ are written as follows:

$$\Phi = \begin{pmatrix} 0 \\ \frac{v+\phi^0}{\sqrt{2}} \end{pmatrix}, \qquad S_F = \begin{pmatrix} \frac{v_s + S_R + i S_I}{\sqrt{2}} \end{pmatrix},$$

where $v_s$ and $v$ stand for the vacuum expectation values (VEVs) of the complex singlet $S_F$ and the SM Higgs doublet, respectively. The scalar potential is expected to be invariant under a flavor symmetry $U(1)_F$, which states that $\Phi \to \Phi$ and $S_F \to e^{i\alpha}S_F$. In general, the FNSM scalar potential allows for the complex VEV $\langle S_F \rangle_0 = \frac{v_s}{\sqrt{2}}e^{i\xi}$. In this work, we consider the special case of $CP$-conserving, i.e., $\xi = 0$.

The $CP$-conserving scalar potential is then given by

$$V_0 = -\frac{1}{2}m_1^2 \Phi^\dagger \Phi - \frac{1}{2}m_2^2 S_F^* S_F + \frac{1}{2}\lambda_1(\Phi^\dagger\Phi)^2 + \lambda_2(S_F^*S_F)^2 + \lambda_3(\Phi^\dagger\Phi)(S_F^*S_F). \quad (2.1)$$

After the spontaneous symmetry breaking by the VEVs of the scalar fields $\Phi$ and $S_F$, the massless Goldstone boson arises. To generate mass to it, we introduce a soft $U(1)_F$ breaking term to the scalar potential in Eq. (2.1):

$$V_{\text{soft}} = -\frac{1}{2}m_3^2(S_F^2 + S_F^{*2}). \quad (2.2)$$

Then, the complete scalar potential is given by

$$V_{\text{FNSM}} = V_0 + V_{\text{soft}}. \quad (2.3)$$

After both the spontaneous symmetry breaking and the minimization conditions on the potential $V_{\text{FNSM}}$ are performed, we identify a mixing between the spin-0 fields via the $\lambda_3$ parameter, which contributes to the mass terms, as follows:

$$m_1^2 = v^2 \lambda_1 + v_s^2 \lambda_3, \quad (2.4)$$

$$m_2^2 = -2m_3^2 + 2v_s^2 \lambda_2 + v^2 \lambda_3. \quad (2.5)$$

Meanwhile, the soft $U(1)_F$ flavor symmetry-breaking term $V_{\text{soft}}$ generates a pseudoscalar flavon mass.

Because all the parameters in the potential in Eq. (2.1) are real, the imaginary and real parts of $V_{\text{FNSM}}$ do not mix. The $CP$-even mass matrix written in the $(\phi_0, S_R)$ basis is given by

$$M_S^2 = \begin{pmatrix} \lambda_1 v^2 & \lambda_3 v v_s \\ \lambda_3 v v_s & 2\lambda_2 v_s^2 \end{pmatrix}.$$

The mass eigenstates are obtained through the standard $2 \times 2$ rotation:

$$\begin{pmatrix} \phi^0 \\ S_R \end{pmatrix} = \begin{pmatrix} \cos\alpha & \sin\alpha \\ -\sin\alpha & \cos\alpha \end{pmatrix} \begin{pmatrix} h \\ H_F \end{pmatrix},$$

where we identify $h$ as the SM-like Higgs boson and $H_F$ is the $CP$-even flavon. The $CP$-odd flavon is associated with the imaginary part of the complex singlet $S_I \equiv A_F$ with mass $M_{A_F} = 2m_3^2$. The physical masses $M_\phi$ ($\phi = h, H_F, A_F$) are related to the parameters of the scalar potential in Eq. (2.1), as follows:

$$\lambda_1 = \frac{(\cos\alpha M_h)^2 + (\sin\alpha M_{H_F})^2}{v^2},$$

$$\lambda_2 = \frac{M_{A_F}^2 + (\cos\alpha M_{H_F})^2 + (\sin\alpha M_h)^2}{2v_s^2},$$

$$\lambda_3 = \frac{\cos\alpha \sin\alpha}{v v_s}(M_{H_F}^2 - M_h^2). \quad (2.6)$$

### B. Yukawa Lagrangian

The $U(1)_F$ invariant Yukawa Lagrangian is given by [23]

$$\mathcal{L}_Y = \rho_{ij}^d \left(\frac{S_F}{\Lambda}\right)^{q_{ij}^d} \bar{Q}_i d_{R_j} \Phi + \rho_{ij}^u \left(\frac{S_F}{\Lambda}\right)^{q_{ij}^u} \bar{Q}_i u_{R_j} \tilde{\Phi}$$
$$+ \rho_{ij}^\ell \left(\frac{S_F}{\Lambda}\right)^{q_{ij}^\ell} \bar{L}_i \ell_{R_j} \Phi + \rho_{ij}^\nu \left(\frac{S_F}{\Lambda}\right)^{q_{ij}^\nu} \bar{L}_i \nu_{R_j} \tilde{\Phi} + \text{H.c.},$$
(2.7)

where $\rho_{ij}^f$ ($f = u, d, \ell, \nu$) are dimensionless parameters of order $\mathcal{O}(1)$, $q_{ij}^f$ is associated with Abelian charges such as they reproduce the observed fermion masses, and $\Lambda$ is identified as the ultraviolet mass scale. The Yukawa couplings from Lagrangian (2.8) can be generated after spontaneously breaking the $U(1)_F$ and EW symmetries. By considering the unitary gauge and making the expansion of the neutral component of the heavy flavon $S_F$ around its VEV $v_s$, one obtains

$$\left(\frac{S_F}{\Lambda}\right)^{q_{ij}} \simeq \left(\frac{v_s}{\sqrt{2}\Lambda}\right)^{q_{ij}} \left[1 + q_{ij}\left(\frac{S_R + iS_I}{v_s}\right)\right]. \quad (2.8)$$

From Eqs. (2.8), (2.7) and after replacing the mass eigenstates, the Yukawa Lagrangian reads

$$\mathcal{L}_Y = \frac{1}{v} [\bar{U} M^u U + \bar{D} M^d D + \bar{L} M^\ell L](c_\alpha h + s_\alpha H_F)$$
$$+ \frac{v}{\sqrt{2} v_s} [\bar{U}_i \tilde{Z}_{ij}^u U_j + \bar{D}_i \tilde{Z}_{ij}^d D_j + \bar{L}_i \tilde{Z}_{ij}^\ell L_j]$$
$$\times (-\sin\alpha h + \cos\alpha H_F + i A_F) + \text{H.c.}, \quad (2.9)$$

where the Higgs-flavon couplings are encapsulated in the $\tilde{Z}_{ij}^f = U_L^f Z_{ij}^f U_L^{f\dagger}$ matrix elements and $M^f$ corresponds to the diagonal fermion mass matrix. The couplings $\tilde{Z}_{ij}^f$ can be written as follows:

$$\tilde{Z}_{ij}^f = \tilde{\rho}_{ij}^f \left(\frac{v_s}{\sqrt{2}\Lambda}\right)^{q_{ij}^f}, \quad (2.10)$$

where $\tilde{\rho}_{ij}^f$ encompasses the diagonalization effects of $Z_{ij}^f$ into $\rho_{ij}^f$. This means that, in general, $\tilde{\rho}_{ij}^f$ is different from $\mathcal{O}(1)$. Equation (2.10) remains nondiagonal even after process of diagonalization of the mass matrices; as a consequence, the FNSM allows for flavor-changing neutral currents. The $q_{ij}^f$ factor can be expressed in terms of the flavor charges of the fermions and the Higgs boson. For quarks, they are given as follows [31]:

$$q_{ij}^d = q_{Q_i} - q_{d_j} - q_H,$$
$$q_{ij}^u = q_{Q_i} - q_{u_j} + q_H, \quad (2.11)$$

where $q_{u_i} = q_{u,c,t}$ and $q_{d_i} = q_{d,s,b}$ stand for the flavor charges of the three generations of quarks singlets, $q_{Q_i}$ corresponds to the flavor charges of the three generations of quark doublets, and $q_H$ denotes the flavor charge of the Higgs boson. In order to obtain the measured quark masses, we set $q_S = +1$, $q_H = 0$ and

$$\begin{pmatrix} q_{Q_1} & q_{Q_2} & q_{Q_3} \\ q_u & q_c & q_t \\ q_d & q_s & q_b \end{pmatrix} = \begin{pmatrix} 3 & 2 & 0 \\ -5 & -2 & 0 \\ -4 & -3 & -3 \end{pmatrix}.$$

The analogous exponents in Eq. (2.10) for the lepton sector read

$$q_{ij}^\ell = q_{L_i} - q_{\ell_j} - q_H,$$
$$q_{ij}^\nu = q_{L_i} - q_{\nu_j} + q_H, \quad (2.12)$$

where $q_{\nu_j} = q_{\nu_e}, q_{\nu_\mu}, q_{\nu_\tau}$ and $q_{\ell_j} = q_e, q_\mu, q_\tau$ are the flavor charges of the three generations of the lepton singlets, while $q_{L_i}$ denotes the flavor charges of the three generations of the lepton doublets. As in the quark sector, we choose the charges to reproduce the lepton masses and mixing patterns, as follows:

$$\begin{pmatrix} q_{L_1} & q_{L_2} & q_{L_3} \\ q_{\nu_e} & q_{\nu_\mu} & q_{\nu_\tau} \\ q_e & q_\mu & q_\tau \end{pmatrix} = \begin{pmatrix} 1 & 0 & 0 \\ -24 & -21 & -20 \\ -8 & -5 & -3 \end{pmatrix}.$$

With the definition $\epsilon = v_s/\sqrt{2}\Lambda$, the fermion mass hierarchy can be expressed as follows:

$$\frac{m_b}{m_t} \approx \epsilon^3, \quad \frac{m_c}{m_t} \approx \epsilon^4, \quad \frac{m_s}{m_t} \approx \epsilon^5, \quad \frac{m_d}{m_t} \approx \epsilon^7, \quad \frac{m_u}{m_t} \approx \epsilon^8$$
$$\frac{m_\tau}{m_t} \approx \epsilon^3, \quad \frac{m_\mu}{m_t} \approx \epsilon^5, \quad \frac{m_e}{m_t} \approx \epsilon^9, \quad m_t \approx \frac{v}{\sqrt{2}}, \quad (2.13)$$

and the Cabibbo-Kobayashi-Maskawa matrix becomes

$$V_{\text{CKM}} \approx \begin{pmatrix} 1 & \epsilon & \epsilon^3 \\ \epsilon & 1 & \epsilon^2 \\ \epsilon^3 & \epsilon^2 & 1 \end{pmatrix}.$$

As far as the $\phi VV$ ($V = W, Z$) interactions are concerned, from the kinetic terms of the Higgs doublet and the complex singlet, we can extract their couplings. Thus, we present the relevant Feynman rules in Table I.

### C. Model parameter space

In order to have realistic predictions, in this section we present a detailed analysis on the model parameter space. We manly focus on the parameters that have a direct impact on the observable studied, i.e., the production cross section of the flavon and its subsequent decay into photons. For this purpose, as we will see in the next section

TABLE I. Tree-level couplings of the SM-like Higgs boson $h$ and the flavons $H_F$ and $A_F$ to fermion and gauge boson pairs in the FNSM. Here, $r_s = v/(\sqrt{2}v_s)$.

| Vertex ($\phi XX$) | Coupling constant ($g_{\phi XX}$) |
|---|---|
| $hf_i\bar{f}_j$ | $\frac{c_\alpha}{v}\tilde{M}^f_{ij} - s_\alpha r_s \tilde{Z}^f_{ij}$ |
| $H_F f_i \bar{f}_j$ | $\frac{s_\alpha}{v}\tilde{M}^f_{ij} + c_\alpha r_s \tilde{Z}^{fij}$ |
| $A_F f_i \bar{f}_j$ | $r_s \tilde{Z}^{fij}$ |
| $hZZ$ | $\frac{gM_Z}{c_W}c_\alpha$ |
| $hWW$ | $gM_W c_\alpha$ |
| $H_F ZZ$ | $g\frac{M_Z}{c_W}s_\alpha$ |
| $H_F WW$ | $gM_W s_\alpha$ |

[Eqs. (3.4)–(3.6)], we require constraining the following free parameters:

(1) Cosine of the mixing angle $\alpha$ and
(2) Vacuum expectation value of the complex singlet $S_F$: $v_s$.

The observables to constrain them include both theoretical and experimental constraints as follows:

(i) Theoretical constraints
   (a) Stability of the scalar potential. The scalar potential in Eq. (2.1) requires absolute stability, i.e., it should not become unbounded from below. The absolute stability demands the following conditions [33]:

$$\lambda_{1,2}(\lambda) > 0, \qquad \lambda_3(\Lambda) + \sqrt{2\lambda_1(\Lambda)\lambda_2(\Lambda)} > 0. \quad (2.14)$$

These potential parameters are evaluted at a scale $\Lambda$ using renormalization group evolution equations (RGE).

   (b) Perturbativity and unitarity constraints.
   We also need to make sure that radiative corrections for the scalar potential remains perturbative at any given energy scale. To ensure this, one must impose upper bounds on the quartic couplings, as follows:

$$|\lambda_{1,2,3}(\Lambda)| \leq 4\pi. \quad (2.15)$$

The quartic couplings are also severely constrained by the unitarity of the S-matrix, which demands that the eigenvalues of it should be less than $8\pi$ [33]. Using the equivalence theorem, the unitary bounds obtained from the S-matrix are given by

$$\lambda_1(\Lambda) \leq 16\pi, \qquad |\lambda_1(\Lambda) + \lambda_2(\Lambda) \pm \sqrt{(\lambda_1(\Lambda) - \lambda_2(\Lambda))^2 + ((2/3)\lambda_3(\Lambda))^2}| \leq 16\pi/3. \quad (2.16)$$

From conditions (2.6), (2.14), and (2.15), we can constrain the mixing angle $\alpha$; the complex singlet VEV $v_s$; and the CP-even and CP-odd flavon masses $M_{H_F}$ and $M_{A_F}$, respectively. For this purpose, we scan the range of values presented in Table II. Figure 1 shows the $\cos\alpha - v_s$ plane, where the blue points correspond to those allowed by all the theoretical constraints, with the unitarity restriction $|\lambda_1(\Lambda) + \lambda_2(\Lambda) + \sqrt{(\lambda_1(\Lambda) - \lambda_2(\Lambda))^2 + ((2/3)\lambda_3(\Lambda))^2}| \leq 16\pi/3$ being the most stringent. According to our analysis, these theoretical constraints impose lower limits on the parameter $v_s$, but not upper limits. Thus, we present the values for $v_s$ in the [0, 200] interval. To avoid dangerous deviations from the $hf\bar{f}$ couplings, $\cos\alpha \approx 1$ is required. Theoretical constraints allow $\cos\alpha = 0$, which contradicts the above. A study is needed to rule out values of $\cos\alpha$ that are inconsistent with experimental observations, as discussed below.

TABLE II. Parameters and ranges scanned.

| Parameter | Range |
|---|---|
| $v_s$(GeV) | [0.001, 5000] |
| $\cos\alpha$ | [−1, 1] |
| $M_{H_F}$(GeV) | [200, 1000] |
| $M_{A_F}$(GeV) | [200, 1000] |

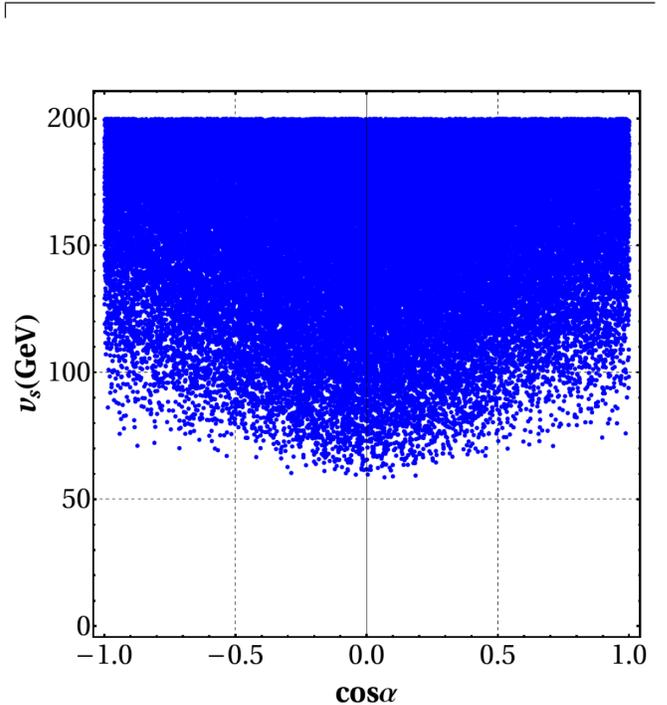

FIG. 1. Parameter space in the $\cos\alpha - v_s$ plane. Blue points stand for these that satisfy all the theoretical constraints.

(ii) Experimental constraints. We also confront the free parameters of the FNSM against experimental measurements. Thus, we analyze the LHC Higgs boson data and its projections for the HL-LHC and HE-LHC [34]. Several experimental upper limits on lepton flavor-violating processes are also included.

(a) LHC Higgs boson data and its projections for the HL-LHC and HE-LHC.

For a decay $S \to X$ or a production process $\sigma(pp \to S)$, the signal strength is defined as [34]

$$\mu_X = \frac{\sigma(pp \to h) \cdot \mathcal{BR}(h \to X)}{\sigma(pp \to h^{SM}) \cdot \mathcal{BR}(h^{SM} \to X)}, \quad (2.17)$$

where $\sigma(pp \to S)$ is the production cross section of $S$, with $S = h, h^{SM}$; here, $h$ is the SM-like Higgs boson coming from an extension of the SM and $h^{SM}$ is the SM Higgs boson; $\mathcal{BR}(S \to X)$ is the branching ratio of the decay $S \to X$, where $X = c\bar{c}, b\bar{b}, \tau^-\tau^+, \mu^-\mu^+, WW^*, ZZ^*, \gamma\gamma$. In our analysis of $\mu_X$, we consider $\tilde{Z}_{bb} = 0.01$ and $\tilde{Z}_{tt} = 0.4$. Such values are well motivated because they simultaneously accommodate all the $\mu_X$'s. In fact, values in the $0.01 \leq \tilde{Z}_{bb} \leq 0.1$ and $0.1 \leq \tilde{Z}_{tt} \leq 1$ intervals have no important impact on the $\mu_X$'s. However, in the case $\tilde{Z}_{bb} \geq 0.1$ and $\tilde{Z}_{tt} \geq 2$, a large reduction of allowed values in the $\cos\alpha - v_s$ plane is found. From Eqs. (2.10) and (2.13), and the corresponding flavor charges in Eq. (2.11), the values of the couplings $\tilde{Z}_{bb}$ and $\tilde{Z}_{tt}$ can be reproduced by assigning $\tilde{\rho}^d_{bb} = 0.41$, $\tilde{\rho}^u_{tt} = 1.63$ and $v_s/\Lambda = 0.41(0.35)$ for $\tilde{Z}_{bb}(\tilde{Z}_{tt})$.

(b) Lepton flavor-violating processes. For this analysis, we study i.) upper limits on the $\mathcal{BR}(\mu \to e\gamma)$ [35], $\mathcal{BR}(\tau \to \mu\gamma)$ [36], $\mathcal{BR}(\tau \to e\gamma)$ [37], ii.) upper limits on $\mathcal{BR}(\mu \to 3e)$, $\mathcal{BR}(\tau \to 3e)$, $\mathcal{BR}(\tau \to 3\mu)$ and $\mathcal{BR}(\mu \to \mu ee)$ [38], iii.) measurements of the $\mathcal{BR}(B_s \to \mu\mu)$ [39] and $\mathcal{BR}(B_d \to \mu\mu)$ [39] and finally iv.) the muon anomalous magnetic moment $\delta a_\mu$ [40]. We find that the processes i.–iii. are not very restrictive in the FNSM. This is mainly because of the choices we made for the matrix elements $\tilde{Z}_{\mu\mu}$ and $\tilde{Z}_{\tau\tau}$, as they play a subtle role in the couplings $\phi\mu\mu$ and $\phi\tau\tau (\phi = h, H_F, A_F)$ (see Table I), which have a significant impact on the observables $\tau \to 3\mu$, $\tau \to \mu\gamma$, $\mu \to e\gamma$. In fact, we consider $\tilde{Z}_{\tau\tau} = 0.2$ and $\tilde{Z}_{\mu\mu} = 10^{-4}$ (hence, a strong hierarchy), otherwise the SM $h\mu\mu$ coupling would be swamped by corrections from the FNSM. As in the case of the elements $\tilde{Z}_{bb}$ and $\tilde{Z}_{tt}$, the values assigned to $\tilde{Z}_{\mu\mu}$ and $\tilde{Z}_{\tau\tau}$ are derived by considering Eqs. (2.10), (2.12), and (2.13), and using the values for $\tilde{\rho}^\ell_{\mu\mu} = 0.16$, $\tilde{\rho}^\ell_{\tau\tau} = 19.42$, and $v_s/\Lambda = 0.32(0.31)$ for $\tilde{Z}_{\mu\mu}$ ($\tilde{Z}_{\tau\tau}$). In contrast, we find that the most stringent constraint comes from $\delta a_\mu$, which imposes an upper limit on the complex singlet VEV $v_s$. However, the situation could change because it is still possible that more precise determinations of the SM hadronic contribution and the experimental measurement would settle the discrepancy in the future without requiring any new physics effects. Thus, in this work, we remain conservative but with an open stance to the above described happening.

Based on the previous sections, we present in Fig. 2 the most restrictive observables of the model parameter space, limitedly to the reduced intervals $-1 \leq \cos\alpha \leq -0.95$ and $0.95 \leq \cos\alpha \leq 1$, since it is the region in which all the analyzed observables converge. The overlapping red and green points indicate the region allowed by all the constraints considered in this work.

It should be noted that quark flavor constraints, namely, $B - \bar{B}$ mixing, $K - \bar{K}$ mixing, and $D - \bar{D}$ mixing, might impose severe restrictions on some model parameters. In Refs. [25,31], this is made evident on the $M_{A_F} - v_s$ plane, which strictly bounds $M_{A_F}$ as a function of $v_s$. We avoid these dangerous bounds via the deletion of the matrix elements $\tilde{Z}_{db}$, $\tilde{Z}_{ds}$, and $\tilde{Z}_{uc}$ (see Table I) for $B - \bar{B}$ mixing, $K - \bar{K}$ mixing, and $D - \bar{D}$ mixing, respectively. We present in Fig. 3 the $M_{H_F} - v_s$ plane for the most restrictive case: $D - \bar{D}$ mixing. We consider the values for $\tilde{Z}_{uc} = 10^{-10}$, $10^{-9}$, and $5 \times 10^{-9}$.

The colored areas are those allowed by $|M^D_{12}|$ [41]:

$$|M^D_{12}| < 7.5 \times 10^{-3} \text{ ps}^{-1}. \quad (2.18)$$

Meanwhile, Fig. 4 shows the $v_s/\Lambda - \tilde{\rho}_{uc}$ plane, where the curves represent values such as they reproduce the assigned values for $\tilde{Z}_{uc}$.

In conclusion, based on the analysis of the theoretical and experimental constraints, we define four benchmark points (BMP) to be used in the simulations in the next section.

(i) BMP1: $v_s = 301$ GeV, $\cos\alpha = 0.998$,
(ii) BMP2: $v_s = 353$ GeV, $\cos\alpha = -0.997$,
(iii) BMP3: $v_s = 888$ GeV, $\cos\alpha = -0.999$, and
(iv) BMP4: $v_s = 191$ GeV, $\cos\alpha = -0.999$.

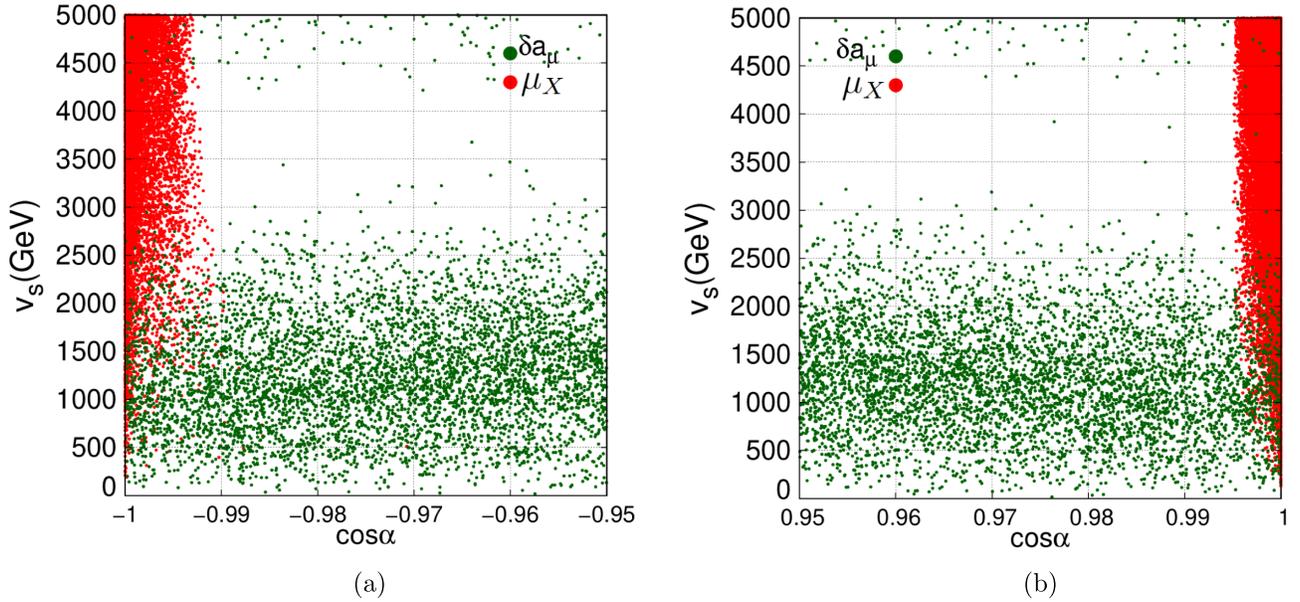

FIG. 2. Interval limited $-1 \leq \cos\alpha \leq -0.95$ (a) and interval limited $0.95 \leq \cos\alpha \leq 1$ (b). Red points correspond to those allowed by all the $\mu_X$'s, while the green points represent those allowed by $\delta a_\mu$. The zone where both regions overlap (red and green points) is the one allowed by all constraints analyzed.

## III. COLLIDER ANALYSIS

This section presents a compressive study on the signal and SM background processes. We also present the strategy for separating one from the other. Explicit analytic expressions for the production mechanism of the flavon and its decay into photons are also presented.

### A. Production and decay of the flavon

We are interested in the production of the $CP$-even flavon $H_F$—the dominant mechanism for producing it is via gluon fusion. This interaction can be extracted through the Lagrangian

$$\mathcal{L}_{\text{eff}} = \frac{1}{v} g_{hgg} h G_{\mu\nu} G^{\mu\nu}, \tag{3.1}$$

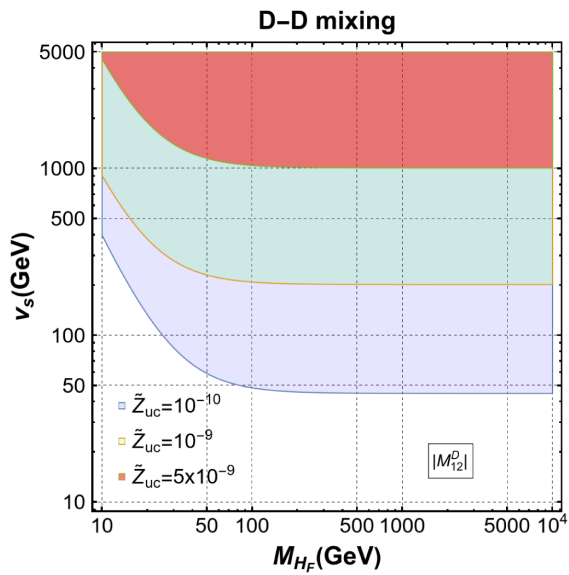

FIG. 3. $M_{H_F} - v_s$ plane showing the region allowed due to flavon contributions to $|M^D_{12}|$. Blue area: $\tilde{Z}_{uc} = 10^{-10}$, green area: $\tilde{Z}_{uc} = 10^{-9}$, red area: $\tilde{Z}_{uc} = 5 \times 10^{-9}$.

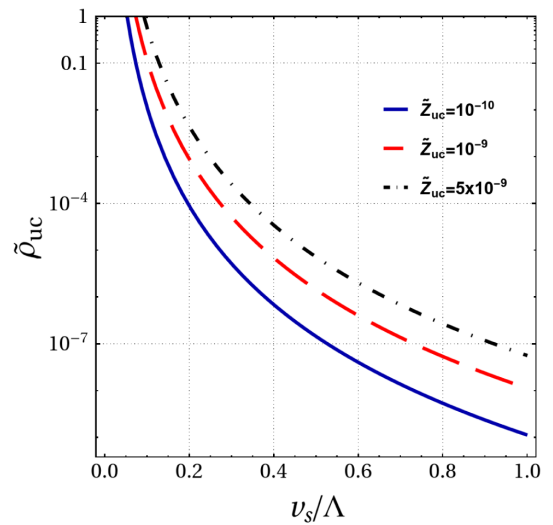

FIG. 4. $\tilde{Z}_{uc}$ as a function of $\tilde{\rho}_{uc}$ and $v_s/\Lambda$, in which $|M^D_{12}|$ is accommodated in the FNSM.

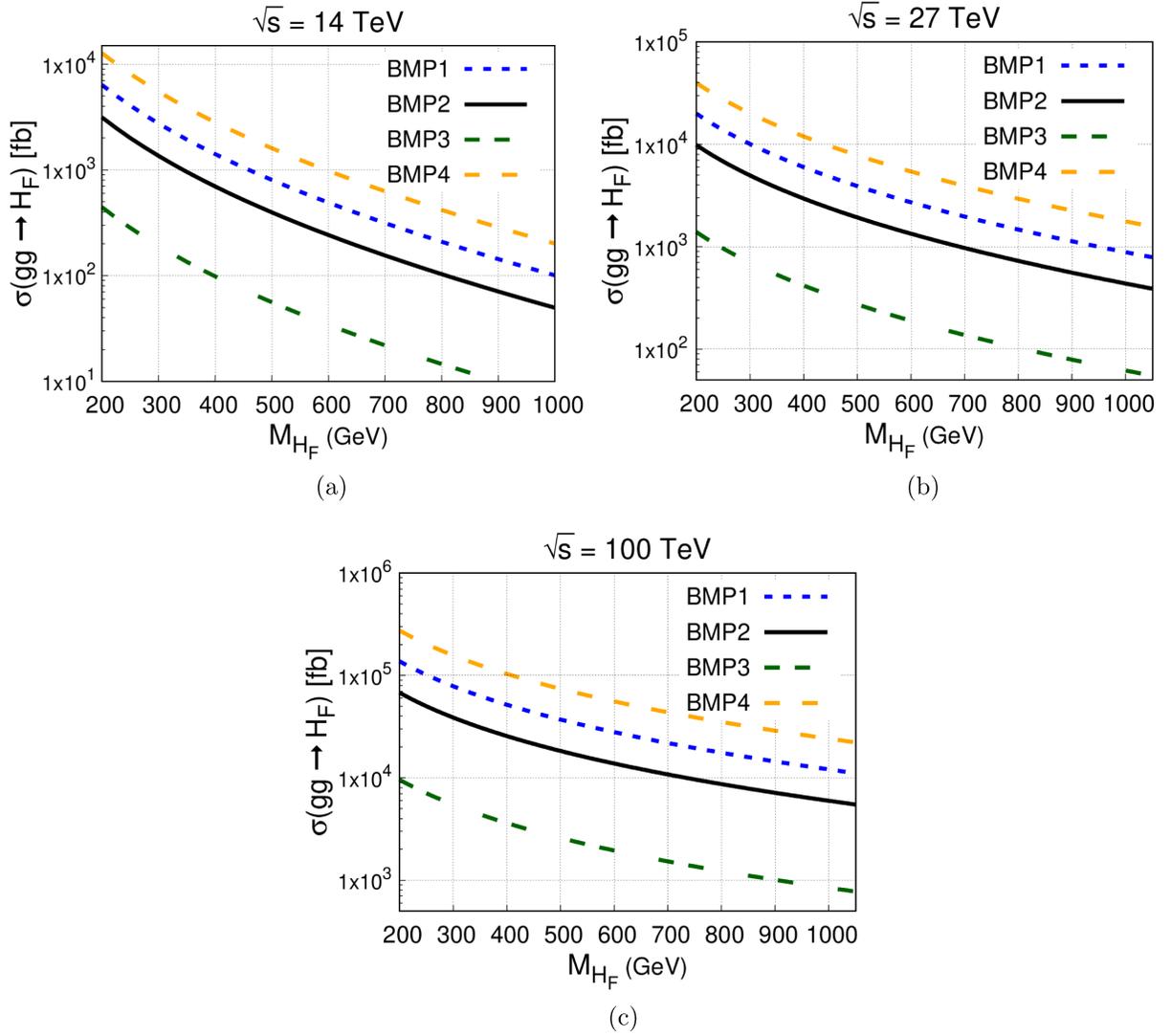

FIG. 5. Production cross section of the flavon via gluon fusion in $pp$ collisions $\sigma(gg \to H_F)$ for the HL-LHC at $\sqrt{s} = 14$ TeV (a), for the HE-LHC at $\sqrt{s} = 27$ TeV (b), and for the FCC-hh at $\sqrt{s} = 100$ TeV (c).

$$g_{Sgg} = -i\frac{\alpha_S}{8\pi}\tau(1 + (1-\tau)f(\tau)) \quad \text{with} \quad \tau = \frac{4M_t^2}{M_h^2}, \quad (3.2)$$

$$f(\tau) = \begin{cases} \left(\sin^{-1}\sqrt{\frac{1}{\tau}}\right)^2, & \tau \geq 1, \\ -\frac{1}{4}\left[\ln\frac{1+\sqrt{1-\tau}}{1-\sqrt{1-\tau}} - i\pi\right]^2 & \tau < 1. \end{cases} \quad (3.3)$$

In FNSM, the $ggh$, $ggH_F$, and $ggA_F$ couplings are given, respectively, by

$$g_{hgg} = \left(\frac{c_\alpha v_s - s_\alpha v}{v_s}\right) g_{Sgg}, \quad (3.4)$$

$$g_{H_F gg} = \left(\frac{c_\alpha v + s_\alpha v_s}{v_s}\right) g_{Sgg}, \quad (3.5)$$

$$g_{A_F gg} = \frac{v}{v_s}(-i\alpha_S/\pi)\tau f(\tau). \quad (3.6)$$

We present in Fig. 5 the production cross section of the flavon $H_F$ as a function of its mass for the colliders: HL-LHC ($\sqrt{s} = 14$ TeV), HE-LHC ($\sqrt{s} = 27$ TeV), and FCC-hh ($\sqrt{s} = 100$ TeV), and the BMPs defined in Sec. II. The most optimistic case for producing flavons is BMP4, which predicts cross sections (at $\sqrt{s} = 14$ TeV), from 200 to $10^4$ fb, corresponding to the range of masses 200–1000 GeV. In contrast, the least favored scenario is BMP3. The cross section of this scenario ranges from 8 fb to 500 fb for $M_{H_F} = 1000, 200$, respectively. These values are expected because the singlet complex VEV $v_s$ suppresses the FNSM correction of the flavon production via gluon fusion, as shown in Eq. (3.5). For $\sqrt{s} = 27, 100$ TeV, the cross sections are up to 1 and 2 orders of magnitude larger than 14 TeV, respectively.

On the other hand, we also need to know the branching ratio of the decay $H_F \to \gamma\gamma$, which can be obtained with the following expression:

$$\Gamma(H_F \to \gamma\gamma) = \frac{\alpha^2 M_{H_F}^3}{1024\pi^3 m_W^2}\left|\sum_s A_s^{H_F\gamma\gamma}(\tau_s)\right|^2, \quad (3.7)$$

where the subscript $s = 0, 1/2, 1$ refers to the spin of the charged particle circulating in the loop, and

$$A_s^{H_F\gamma\gamma} = \begin{cases} \sum_f \frac{2m_W g_{H_F f_i \bar{f}_i} N_c Q_f^2}{m_f}[-2\tau_s(1 + (1-\tau_s)f(\tau_s))] & \text{for } s = \frac{1}{2}, \\ g_{H_F WW}[2 + 3\tau_W + 3\tau_W(2-\tau_W)f(\tau_W)] & \text{for } s = 1, \end{cases} \quad (3.8)$$

where

$$f(x) = \begin{cases} \arcsin^2\left(\frac{1}{\sqrt{x}}\right), & x \geq 1, \\ -\frac{1}{4}\left[\log\left(\frac{1+\sqrt{1-x}}{1-\sqrt{1-x}}\right) - i\pi\right]^2, & x < 1, \end{cases} \quad (3.9)$$

and $N_c = 1, 3$ for leptons and quarks, respectively, $\tau_a = 4M_a^2/M_{H_F}^2$. The couplings $g_{H_F f_i \bar{f}_i}$ and $g_{H_F WW}$ are given in Table I. Figure 6 shows the $\mathcal{BR}(H_F \to \gamma\gamma)$ as a function of the flavon mass $M_{H_F}$ for the BMPs defined in Sec. II.

According to our analysis of the FNSM parameter space, branching ratios from $\mathcal{O}(10^{-6})$ to $\mathcal{O}(10^{-4})$ can be obtained, for BMP2 ($M_{H_F} \sim 400$ GeV) and BMP1 ($M_{H_F} \sim 250$ GeV), respectively; while for BMP3 and BMP4, we obtain similar branching ratios for the range $320 \leq M_{H_F} \leq 1000$ GeV. The discontinuous behavior in all BMPs is due to several factors, the most distinctive being the emergence or suppression of different flavon decay channels. In particular, once $M_{H_F} \sim 250$ GeV, there is an inflection point associated with the emergence of the $H_F \to hh$ di-Higgs channel. For $M_{H_F} > 1000$ GeV, the $\mathcal{BR}(H_F \to \gamma\gamma)$ converges to a value of order $10^{-5}$. The values of the branching ratios are relatively high due to the BMPs defined for our study. From Eq. (3.8) and Table I, we notice that $\mathcal{BR}(H_F \to \gamma\gamma) \sim 1/v_s$, with $v_s$ being of order $10^2$ in the BMPs defined, it suppresses the $\mathcal{BR}(H_F \to \gamma\gamma)$ by a factor of $10^{-2}$ with respect to the $\mathcal{BR}(h \to \gamma\gamma)$ of the SM, which is of order $10^{-3}$. Thus, the $\mathcal{BR}(H_F \to \gamma\gamma)$ shown in Fig. 6 are reasonable and motivate their possible experimental scrutiny.

We also present in Fig. 7 a comparison of the diphoton channel with the dominant flavon decays. According to the BMPs defined above, the flavon particle behaves as quasifermionphobic, with the $t\bar{t}$ channel being the only one to stand out among all the fermion channels.

Concerning our computation scheme, we first implement the model via FeynRules [42] to obtain the UFO files [43] for MadGraph5 [44], which is interfaced with Pythia8 [45] for parton showering, and Delphes [46] for detector response. The Delphes cards used in the simulations correspond to the three colliders considered in this work, namely, `delphes_card_HLLHC.tcl` and `FCChh.tcl`.

### B. Signal and background

(i) *Signal:* We search for a final state $\gamma\gamma$ generated by the decay $H_F \to \gamma\gamma$, where the flavon $H_F$ is produced by gluon fusion. The analysis is performed in the mass range of 200–1000 GeV. We present in Fig. 8 the Feynman diagram of the signal, while Fig. 9 shows the production cross section $\sigma(gg \to H_F \to \gamma\gamma)$ (left vertical axis) and the number of events produced (right vertical axis) as a function of $M_{H_F}$. The numerical cross sections of the signal for the BMPs defined above are presented in Table III.

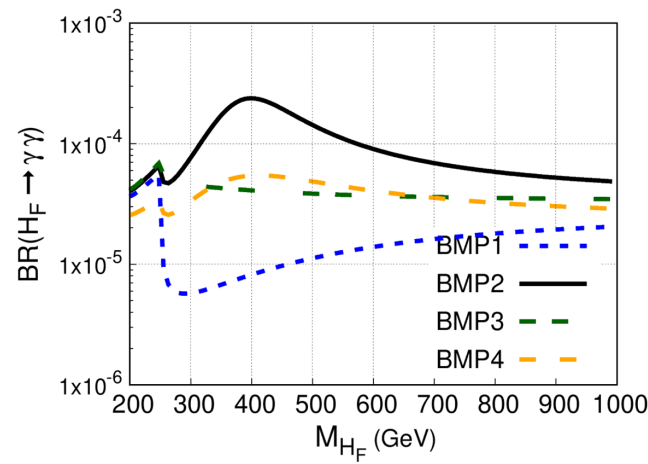

FIG. 6. Branching ratio $\mathcal{BR}(H_F \to \gamma\gamma)$ as a function of the flavon mass $M_{H_F}$.

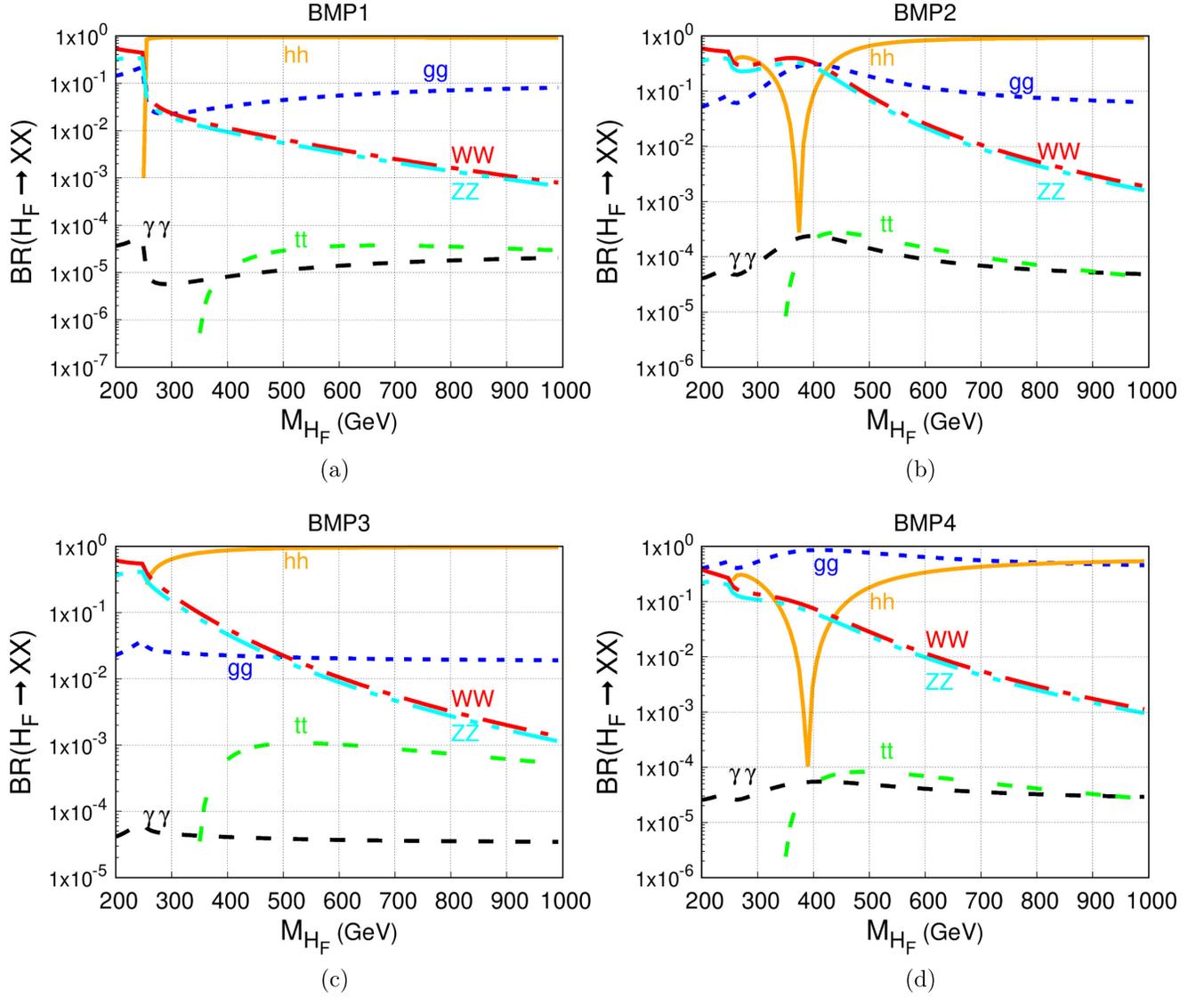

FIG. 7. Branching ratios of the quasifermionphobic flavon into several decay channels for (a) BMP1, (b) BMP2, (c) BMP3, and (d) BMP4.

(ii) *Background:* The dominant irreducible background is the SM diphoton production ($\gamma\gamma$); contributions also come from $\gamma + j$ and $j + j$ production with one or two jets misidentified as photons, and from the Drell-Yan process. The numerical cross sections of the background processes are presented in Table III.

### C. Event selection

The identification of the signal depends mainly on the true $p_T^\gamma$ of the photons. The photon selection efficiency as a function of the $p_T^\gamma$ of the true photon is parametrized by [47]

$$\epsilon(p_T^\gamma) = 0.76 - 1.98 \times e^{-\frac{p_T^\gamma}{16.1 \text{ GeV}}}. \quad (3.10)$$

On the other hand, the rate of jets passing the photon identification and isolation requirements can be identified as fake photons. The rate is parametrized as a function of the true jet $p_T^j$ via [47]

$$\epsilon(p_T^j) = 9.3 \times 10^{-3} \times e^{-\frac{p_T^j}{27.5 \text{ GeV}}}. \quad (3.11)$$

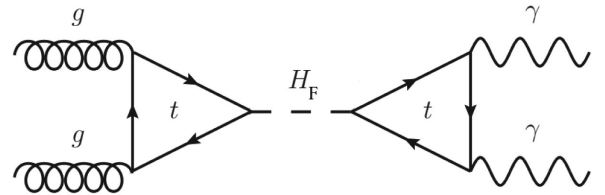

FIG. 8. Feynman diagram of the signal $gg \to H_F \to \gamma\gamma$.

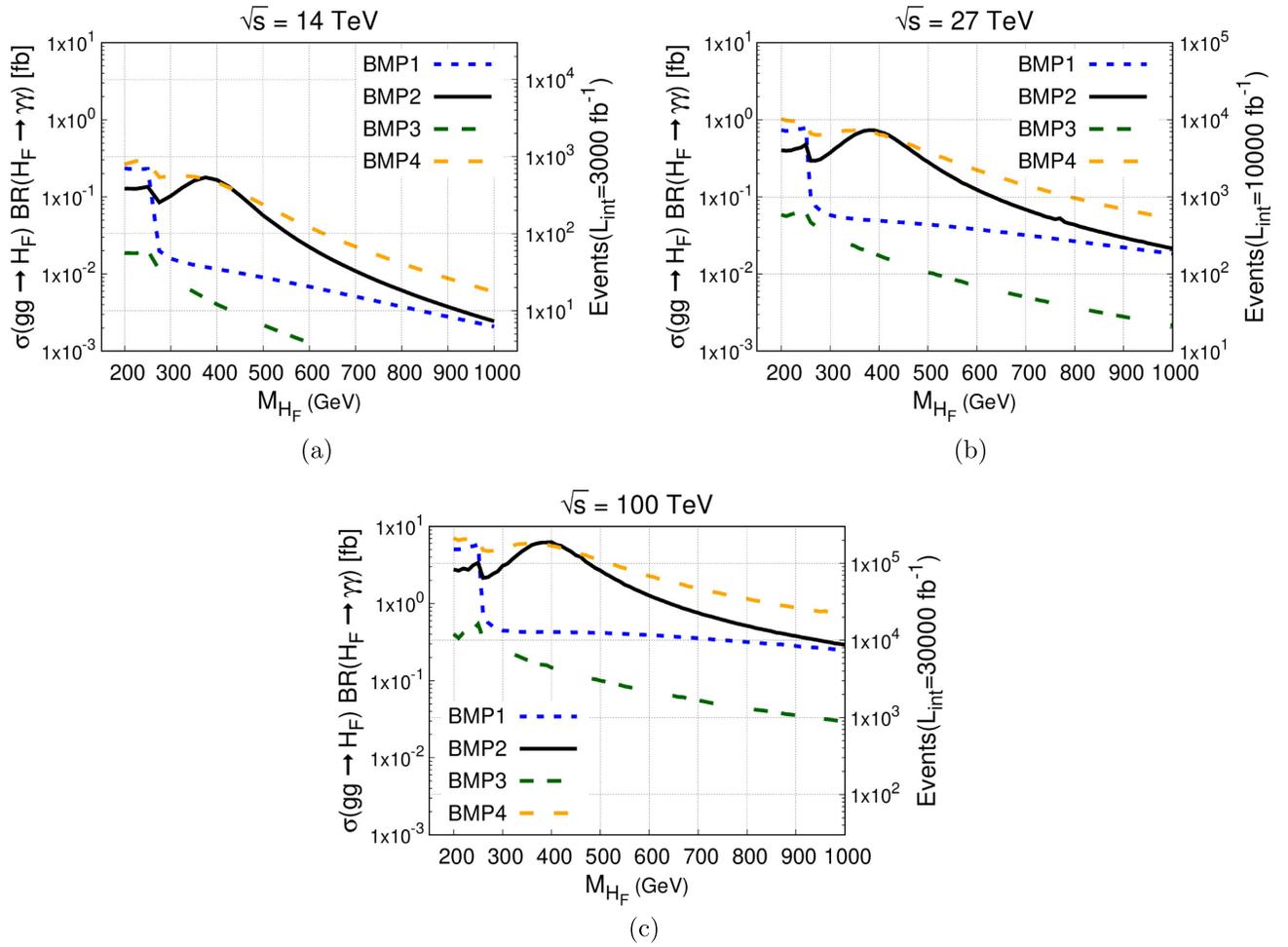

FIG. 9. On the left vertical axis: production cross section $\sigma(pp \to H_F \to \gamma\gamma)$ as a function of the flavon mass $M_{H_F}$ for (a) HL-LHC, (b) HE-LHC and (c) FCC-hh. On the right vertical axis: Number of events produced as a function of the flavon mass $M_{H_F}$.

TABLE III. Cross sections of the signal for $M_{H_F} = 400$ GeV and the SM background processes (here $\ell = e, \mu$). For signal, we consider a cut on the transverse momentum of the photons: $P_T^\gamma > 60$ GeV. This value corresponds to an efficiency of the photon identification of $\epsilon(p_T^\gamma) \sim 0.71$.

| Signal | Cross section (pb) $\sqrt{s} = 14$ TeV | Cross section (pb) $\sqrt{s} = 27$ TeV | Cross section (pb) $\sqrt{s} = 100$ TeV |
|---|---|---|---|
| BMP1 | $1.1 \times 10^{-5}$ | $4.5 \times 10^{-5}$ | $4.6 \times 10^{-4}$ |
| BMP2 | $1.5 \times 10^{-4}$ | $5.1 \times 10^{-4}$ | $3.7 \times 10^{-3}$ |
| BMP3 | $3.9 \times 10^{-6}$ | $1.4 \times 10^{-5}$ | $9.9 \times 10^{-5}$ |
| BMP4 | $1.4 \times 10^{-4}$ | $5.1 \times 10^{-4}$ | $4.7 \times 10^{-3}$ |

| Background | Cross section (pb) $\sqrt{s} = 14$ TeV | Cross section (pb) $\sqrt{s} = 27$ TeV | Cross section (pb) $\sqrt{s} = 100$ TeV |
|---|---|---|---|
| $\gamma\gamma$ | 153 | 239 | 569 |
| $jj$ | $p_T^j > 150$  $2.5 \times 10^5$ | $p_T^j > 300$  $4.5 \times 10^4$ | $p_T^j > 500$  $7 \times 10^4$ |
| $\gamma j$ | $p_T^j > 150$  127 | $p_T^j > 300$  27 | $p_T^j > 500$  29 |
| $pp \to Z(\gamma^*) \to \ell\ell$ | $1.8 \times 10^3$ | $3.2 \times 10^3$ | $9.5 \times 10^3$ |

TABLE IV. List of the variables used to train and test the signal and background events.

| Rank | Variable | Description |
|---|---|---|
| 1 | $p_T^{\gamma_1}$ | Photon with the largest transverse momentum. |
| 2 | $M_{\text{inv}}(\gamma\gamma)$ | Invariant mass |
| 3 | $N(j)$ | Number of jets |
| 4 | $\Delta R$ | The $R$ separation between the photons |
| 5 | $p_T^{\gamma_2}$ | Transverse momentum of the subleading photon |
| 6 | $\eta_1$ | Pseudorapidity of the leading photon |
| 7 | $\eta_2$ | Pseudorapidity of the subleading photon |
| 8 | $\phi(\gamma_1)$ | Azimuth angle of the leading photon |
| 9 | $\phi(\gamma_2)$ | Azimuth angle of the subleading photon |

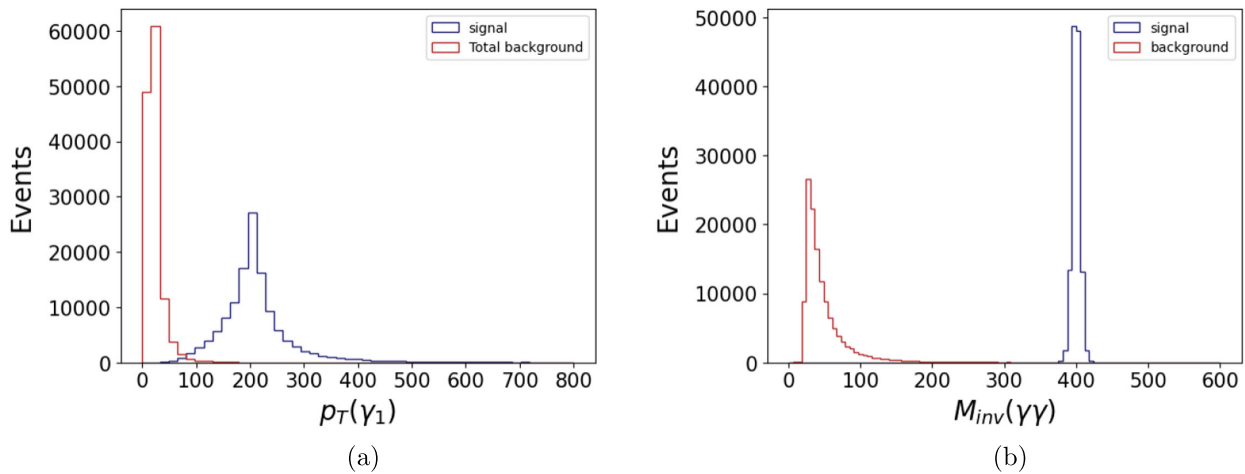

FIG. 10. Transverse momentum of the leading photon $p_T^{\gamma_1}$ (a) and invariant mass $M_{\text{inv}}(\gamma\gamma)$ (b) for $M_{H_F} = 400$ GeV.

The strategy to isolate the signal from the background is to impose the criteria in Eqs. (3.10) and (3.11). For this purpose, we take into account the rate of fake photons through Monte Carlo simulations of dijet and $\gamma j$ events. Thus, we impose a cut on the transverse moment of the jets $p_T^j$ such that the fake photon rate is of the order of $\mathcal{O}(10^{-6})$ and the efficiency of the photon identification to be $\epsilon(p_T^\gamma) \sim 0.71$. The event selection (signal and background) requires that at least one photon per event meets the criteria $\epsilon(p_T^\gamma)$. This effectively suppresses the majority of the events arising from the dijet final state. The photons must be reconstructed within the fiducial region, $|\eta^\gamma| < 2.5$, excluding the barrel-endcap transition region, $1.44 < |\eta^\gamma| < 1.57$. A $p_T^\gamma$ threshold of $M_{\gamma\gamma}/3$ ($M_{\gamma\gamma}/4$) is applied to the leading (subleading) photon in $p_T^\gamma$, where $M_{\gamma\gamma}$ is the diphoton invariant mass. Jet selection criteria are applied to the two jets of largest $p_T^j$ in the event within $|\eta^j| < 4.7$.

In addition, we perform a boosted decision trees (BDT) training [48] by using variables related to the kinematics of the final state. Table IV shows the variables used to train and test the signal and background events. According to our analysis, the most discriminating observables are the transverse momentum of the leading photon $p_T^{\gamma_1}$ and the invariant mass $M_{\text{inv}}(\gamma\gamma)$. We show in Fig. 10 these distributions for $M_{H_F} = 400$ GeV.[1] Meanwhile, Fig. 11 presents the discriminant for the signal and background. The goodness of fit is checked via the Kolmogorov-Smirnov (KS) test. Our analysis shows that the KS value is within the permissible [0, 1] interval, namely, 0.29 and 0.91 for the background and the signal, respectively. The relevant hyperparameters for the BDT training are as follows: number of trees (NTree) = 110, maximum depth of the decision tree (MaxDepth) = 4, maximum number of leaves (MaxLeaves) = 14. The training is performed using the Monte Carlo-simulated samples. These signal and background samples are scaled to the expected number of candidates, which is calculated based on the integrated luminosity and cross sections. The BDT selection is optimized individually for each flavon mass $M_{H_F}$ to maximize the figure of merit, i.e., the *signal significance*, defined as $S/\sqrt{S+B}$, where $S$ and $B$ represent the number of signal and background candidates, respectively, in the signal region after applying the selection criteria.

---

[1]It is important to highlight that these distributions were generated with the same number of signal and background events which were used for training and testing of BDT.

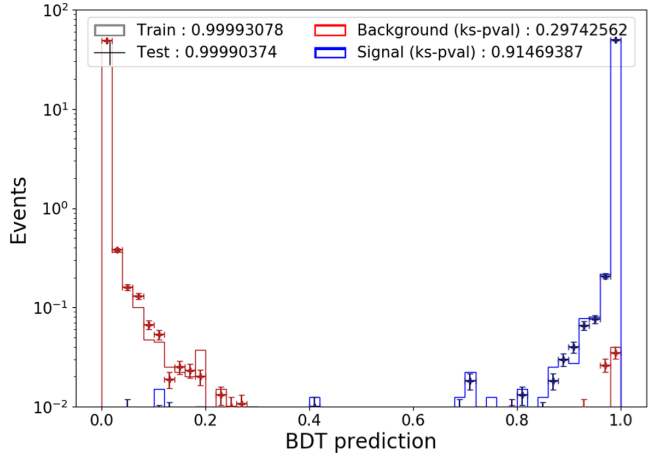

FIG. 11. Normalized plot of the discriminant for signal and background data. This plot corresponds to BMP4 and $M_{H_F} = 400$ GeV.

Figure 12 presents contour plots of the signal significance as a function of the flavon mass $M_{H_F}$ and the integrated luminosity for the BMP2. In particular, we obtain signal significances at the level of $5\sigma$ for $350 \lesssim M_{H_F} \lesssim 450$ GeV and $200 \leq M_{H_F} \leq 1000$ GeV for the HE-LHC and FCC-hh, respectively. The least-favored scenario is for the HL-LHC, which does not provide the ease of detection of the flavon. Similarly, Fig 13 presents the signal significance for the benchmark point BMP4 with similar results for the HL-LHC. However, this scenario provides a range of masses that could be detected slightly greater than the previous case. Specifically, the HE-LHC offers the possibility of detection of the flavon in the range of masses 200–450 GeV. More encouraging results appear at the FCC-hh, achieving a potential discovery in a wider range of masses, covering the entire mass spectrum studied in this paper. In contrast, for the least hopeful case, BMP3, we found a maximum significance of $1.05\sigma$ for the FCC-hh by considering its final integrated luminosity and $M_{H_F} = 200$ GeV. As far as BMP1 is concerned, we find results that approach an intermediate case of BMPs 2 and 3 for $400 \lesssim M_{H_F} \lesssim 600$ GeV and significances similar to BMP2 for $750 \lesssim M_{H_F} \lesssim 1000$. The most relevant signal significances for BMP1, even at the level of the most optimistic BMPs (2 and 4), lie in the mass interval $200 \leq M_{H_F} \leq 250$.

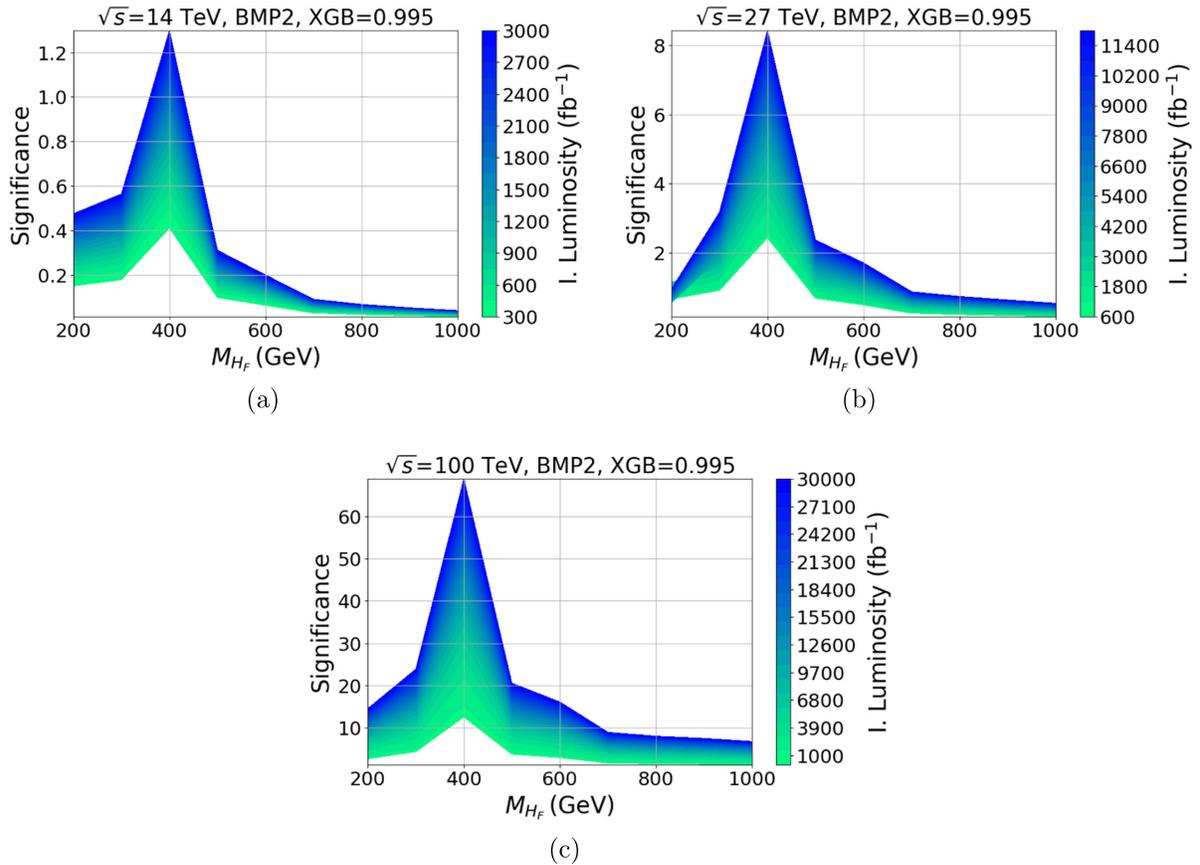

FIG. 12. Significance as a function of the flavon mass $M_{H_F}$ and the integrated luminosity, for the BMP2 at the HL-LHC (a), HE-LHC (b), and FCC-hh (c). In all the cases, we impose a cut on the BDT prediction of XGB = 0.995.

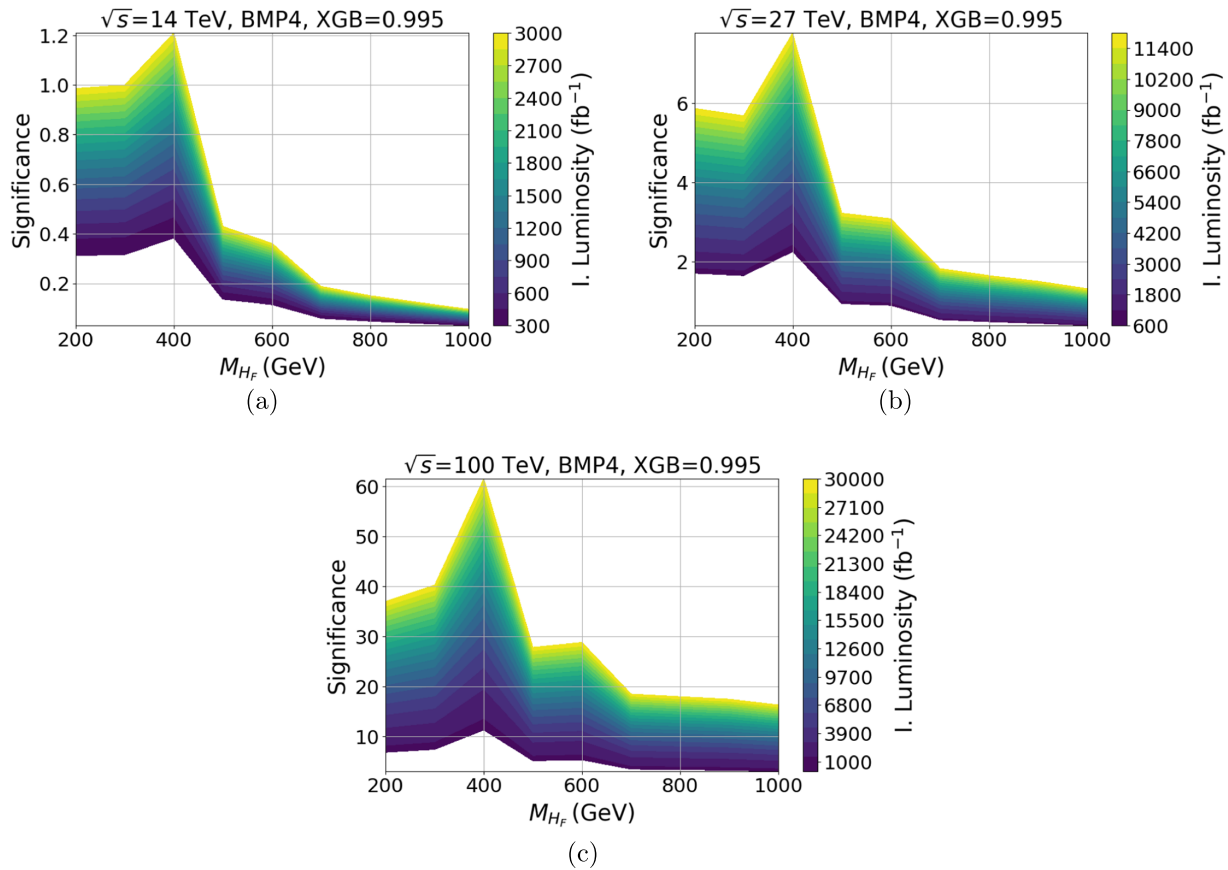

FIG. 13. Significance as a function of the flavon mass $M_{H_F}$ and the integrated luminosity, for the BMP4 at the HL-LHC (a), HE-LHC (b), and FCC-hh (c). In all the cases, we impose a cut on the BDT prediction of XGB = 0.995.

## IV. CONCLUSIONS

In this paper, we have studied the production of the so-called flavon $H_F$, with its subsequent decay into two photons (signal). The flavon is a hypothetical particle predicted in a model that invokes the Froggatt-Nielsen mechanism, which is the theoretical framework considered in this paper. The mechanism of production of the flavon is via proton-proton collisions at super hadron colliders, namely the HL-LHC, HE-LHC and the FCC-hh. To give realistic predictions, we test the free parameters of the model against theoretical and experimental constraints and then we define benchmark points to perform Monte Carlo simulations. The most severe constraint on the cosine of the mixing angle ($\cos \alpha \sim -1$ or $\sim 1$) that mixes the neutral and real parts of the Higgs doublet and the complex singlet with the physical fields $h$ and $H_F$ comes from the LHC Higgs boson data and their projections for the HE-LHC. This is expected in order to avoid dangerous corrections to the Higgs boson couplings to fermions and gauge bosons. As far as the VEV of the complex singlet $v_s$ is concerned, the muon anomalous magnetic moment is the most restrictive observable on $v_s$, allowing $v_s \lesssim 3$ TeV. However, the situation could change because it is still possible that more precise determinations of the SM hadronic contribution and the experimental measurement would settle the discrepancy in the future without requiring any new physics effects. Thus, in this work we remain conservative but with an open stance to the above described happening.

On the other hand, based on the defined BMPs, and using events simulated for the signal and its SM background, we perform an analysis with machine learning via the boosted decision trees method to isolate the signal from the background. We find that the nearest evidence could emerge at the HE-LHC with a signal significance of $5\sigma$ for integrated luminosities in the range 5-12 ab$^{-1}$ and flavon masses between 350–450 GeV (200–450) for BMP2 (BMP4). These predictions could be corroborated in the FCC-hh. Furthermore, this collider would have the capability to search for broader range of masses, covering the entire interval studied in this work. Thus, we predict signal significances at the level of $5\sigma$ for the range $200 \leq M_{H_F} \leq 1000$ GeV. Even more, projecting our results, the FCC-hh might be able to search for flavon masses as high as 5 TeV.

## ACKNOWLEDGMENTS

The work of M. A. A.-U. and T. V.-P. is supported by "Estancias Posdoctorales por México (SECIHTI)" and "Sistema Nacional de Investigadores" (SNII-SECIHTI). T. V.-P. acknowledges support from the UNAM Project No. PAPIIT IN111224 and the Secretaría de Ciencia, Humanidades, Tecnología e Innovación de México (SECIHTI) Project No. CBF2023-2024-548.

## DATA AVAILABILITY

The data that support the findings of this article are openly available at [49].